\documentstyle[11pt,appb,epsf]{article} 
\begin{document}
%\eqsec  % uncomment this line to get equations numbered by (sec.num)
\title{STATUS OF EFFECTIVE FIELD THEORY OF NN SCATTERING}
% you can use \\ to break lines

\author{SILAS R.~BEANE
\address{Department of Physics, University of Maryland\\
College Park, MD 20742-4111}}
\maketitle
\begin{abstract}
I review recent progress in developing a systematic power counting
scheme for scattering processes involving more than one nucleon.
\end{abstract}
%\PACS{25.40.ve}

%%%%%%%%%%%%%%%%%%%%%%%%%%%%%%%%%%%%%%%%%%%%%%%%%%%%%%%%%%%%%%%%%%%%%%%  
\section{Why effective field theory?}

There exist many nucleon-nucleon potentials which reproduce phase
shifts and nuclear properties with remarkable accuracy (an extensive
reference list can be found in Ref.~\cite{Be98a}). Three fundamental
features are shared by these potential models: (i) pions are important
at long distances, (ii) there is a source of intermediate-range
attraction, and (iii) there is a source of short-distance
repulsion. However, in general, distinct physical mechanisms in these
models account for the same feature of the nuclear force.  Agreement
with experiment is maintained in spite of these differences because of
the large number of fit parameters.

Systematic approaches to the scattering of strongly interacting
particles, such as chiral perturbation theory, are based on the ideas
of effective field theory (EFT). The fundamental premise of EFT is
that when a system is probed at momentum $k \ll M$, details of the
dynamics at scale $M$ are unimportant. What is important at low
energies is the physics that can be captured in operators of
increasing dimensionality which take the form of a power-series in
$k/M$~\cite{Ka95,mano}.  It is entirely possible that EFT fits to
phase shifts will ultimately not be as good as those produced by
conventional $NN$ potentials with the same number of parameters. So
then, what can be gained from such an enterprise?

Consider the following questions: Is it possible to account for short
distance physics at low energies systematically, using power counting
arguments?  What is the minimal set of parameters required to describe
data at low energies? Or rather, what is the minimal short distance
physics required?  Can we fit some processes to experiment and use
that information to predict other processes? For instance, one would
like to relate $NN$ scattering systematically to scattering processes
with more nucleons, such as $NNN$ scattering, and to make predictions
for processes involving electromagnetic and pionic probes of
few-nucleon systems.  The underlying theory, QCD, has one scale,
$\Lambda_{QCD}$.  Why are characteristic nuclear binding energies $\ll
\Lambda_{QCD}$?  These are the sort of questions that EFT can help
answer.  In what follows I will review recent progress in answering
some of these questions.

%%%%%%%%%%%%%%%%%%%%%%%%%%%%%%%%%%%%%%%%%%%%%%%%%%%%%%%%%%%%%%%%%%%%%%%
\section{The Weinberg program}

\begin{figure}[t]
\begin{center}
\leavevmode
\epsfxsize=12.0cm
\epsfysize=2.0cm
\epsfverbosetrue
\epsffile{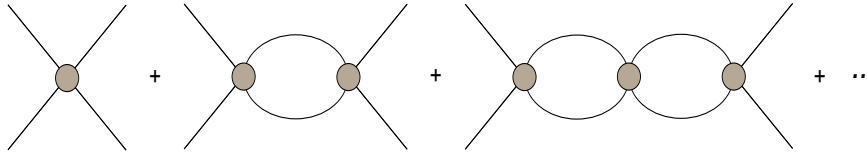}
\end{center}
\caption{The diagrammatic solution of the Schr\"odinger equation with 
the effective $NN$ potential represented by the shaded blob. }
\label{fig1}
\end{figure}

Naive application of EFT ideas to nuclear physics immediately suggests
a puzzle.  In nuclear physics there are bound states whose energy is
unnaturally small on the scale of hadronic physics. In order to
generate such bound states within a ``natural'' theory it is clear
that one must sum some operators to all orders.  Weinberg
proposed~\cite{We90,We91} implementing the EFT program in nuclear
physics by applying the power counting arguments of chiral
perturbation theory to an $n$-nucleon effective potential rather than
directly to the S-matrix. Only $n$-nucleon irreducible graphs should
be included in the $n$-nucleon effective potential. The potential
obtained in this way is then to be inserted into a Schr\"odinger
equation and iterated to all orders.  See Fig.~\ref{fig1}.  There will
of course be unknown coefficients in the effective potential, but
these can be fit to experimental data as in ordinary chiral
perturbation theory~\cite{Or96,Ka96,Sc97,Le97,park}.  Perhaps
the most powerful result to emerge from Weinberg's power counting is
the hierarchy of $n$-body forces (e.g. three-body forces are
small)~\cite{We90,We91,Ubi94,Fr96}.

The regularization and renormalization of the potential is
straightforward in Weinberg's scheme. However, Weinberg did not
specify how to regularize and renormalize the Schr\"odinger
equation. As we will see, an understanding of regularization and
renormalization appears to be crucial in the identification of a
consistent power counting scheme.

%%%%%%%%%%%%%%%%%%%%%%%%%%%%%%%%%%%%%%%%%%%%%%%%%%%%%%%%%%%%%%%%%%%%%%%
\section{Dissecting the Weinberg program -- the pionless EFT}

In Ref.~\cite{Ka96} Kaplan, Savage and Wise (KSW) considered $NN$
scattering in the ${}^1S_0$ $(np)$ channel at momentum scales $k \ll
m_\pi$.  The EFT at these scales involves only nucleons since the pion
is heavy and may therefore be ``integrated out''.  The effective
Lagrangian thus consists of contact operators of increasing
dimensionality constrained by spin and isospin. This EFT is useful
because scattering amplitudes can be calculated analytically. It
therefore allows one to address issues of principle in EFT for $NN$
scattering.

The most general effective Lagrangian consistent with spin and
isospin, including only operators relevant to ${}^1S_0$ scattering is

\begin{equation}
{\cal L}=N^\dagger i \partial_t N - N^\dagger \frac{\nabla^2}{2 M} N
- \frac{1}{2} C (N^\dagger N)^2\\ 
-\frac{1}{2} C_2 (N^\dagger \nabla^2 N) (N^\dagger N) + h.c. + \ldots.
\label{eq:lag}
\end{equation}
It is important to realize that all of the coefficients in the
effective theory, $C, C_2,...$, are renormalization scheme dependent.
This means that power counting will necessarily look different in
different schemes. It is clearly fruitful to choose a scheme which
maintains the power counting hierarchy of operators; although
ultimately the scattering amplitude which is calculated to a given
order in the EFT is scheme independent, the power counting is
transparent in some schemes while requiring counterintuitive
cancelations in others~\cite{mano}. For instance, in a perturbative
EFT expansion --such as Fermi theory-- power counting is transparent
in dimensional regularization ($DR$) with minimal subtraction ($MS$),
while somewhat mysterious using a cut-off~\cite{mano}.

\begin{figure}[t]
\begin{center}
\leavevmode
\epsfxsize=10.0cm
\epsfysize=8.0cm
\epsfverbosetrue
\epsffile{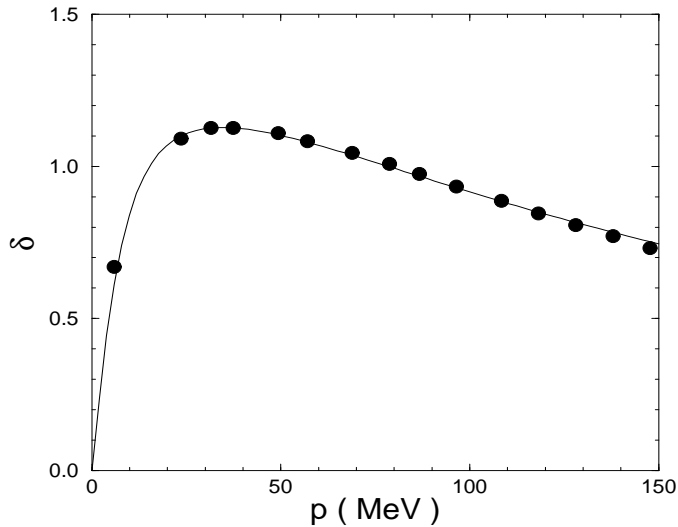}
\end{center}
\caption{The effective range expansion with the extracted phase-shift
data up to center-of-mass momenta of order $M_\pi$.}
\label{fig2}
\end{figure}

Ultimately what one would like to reproduce in the $NN$ EFT is the
effective range expansion, written here as

\begin{equation}
\frac{1}{T^{\rm on}(k)}=-\frac{M}{4 \pi}\left[ -\frac{1}{a} +
\frac{1}{2} r_e {k^2} +O(k^4)- i k \right],
\label{eq:cutoff2}
\end{equation}
where $T(k)$ is the scattering amplitude, $a$ is the scattering length
and $r_e$ is the effective range.  Experiment determines (in the
${}^1S_0$ $(np)$ channel) $a=-23.714\pm 0.013\, {\rm fm}$ and ${r_e}=2.73\pm
0.03\, {\rm fm}$.  The extremely large value of the scattering length
implies that there is a virtual bound state in this channel very near
zero energy. While the value of $r_e$ is consistent with what one
might expect for a natural theory where pions dominate the low-energy
physics ($r_e \sim 1/M_\pi$), the value of $a$ is far from being
natural ($a \gg 1/M_\pi$).  As seen in Fig.~\ref{fig2} this scattering
amplitude (neglecting $O(k^4)$ terms) compares favorably with data up
to up to center-of-mass momenta of order $M_\pi$. Phase shift data of
Fig.~\ref{fig2} are taken from Ref.~\cite{St93}.

I will proceed in the spirit of Weinberg power counting. The effective
potential in the pionless EFT is simply the sum of all tree graphs
extracted from the lagrangian of Eq.~(\ref{eq:lag}). It is
straightforward to find the ``second-order'' potential:

\begin{equation}
V^{(2)}(p',p)=C + C_2(p^2 + p'^2).
\label{eq:V2}
\end{equation}
The Schr\"odinger equation iterates this potential to all orders (see
Fig.~\ref{fig1}).  The divergences get worse as one goes to higher
order in the potential.  All divergences are of power-law
type. Therefore $DR$ with $MS$ has the effect of unitarizing the
scattering amplitude with the potential from
Eq.~(\ref{eq:V2})~\cite{Be98a,Le97}. The problem is that the resulting
scattering amplitude only matches to the effective range expansion for
momenta $k\ll 1/\sqrt{a{r_e}}$.  Given our working assumption that all
renormalization schemes give equivalent results but generally have
different power counting, this means that $DR$ with $MS$ is not
particularly well suited to the problem since higher order operators
must be highly correlated in this scheme in order to ensure that the
EFT matches to the effective range expansion~\cite{Ka96}.  In
Ref.~\cite{Ka97} a novel way of reproducing the effective range
expansion within the $DR$ with $MS$ scheme was proposed. The main idea
is that the effective range expansion can be viewed as arising from
the exchange of a di-baryon field (transvestite in the vernacular)
which is included in the EFT as a fundamental field and ``dressed''
via its interactions with the nucleons.

Following the work of Ref.~\cite{Ka96} many authors argued that the
the pathological features of $DR$ with $MS$ are a good reason to work
with a cut-off EFT~\cite{Le97,tom,PC97,Ri97}. The problem is that,
unless one is willing to carry out all analysis numerically, not much
insight is gained into power counting; the unpleasant features that
one has in cut-off Fermi theory are present in $NN$ scattering with a
vengeance. There are other pathologies as well which force the cut-off
to be very low~\cite{tom,PC97,Sc97,Be98a,Ri97}, unless the bare
coefficients in the lagrangian are chosen to be imaginary. We will
return to the issue of cut-off EFT below.

%%%%%%%%%%%%%%%%%%%%%%%%%%%%%%%%%%%%%%%%%%%%%%%%%%%%%%%%%%%%%%%%%%%%%%%
\section{Resolution}

The physical scattering amplitude that is generated when an effective
potential is iterated in the Schr\"odinger equation is exactly unitary
(like Eq.~(\ref{eq:cutoff2})) and therefore necessarily contains
arbitrarily high powers in momentum. This occurs regardless of the
order to which one is working in the momentum expansion of the
potential $V$. Therefore, the scattering amplitude thus obtained
samples arbitrarily short-distance scales. Such a scattering amplitude
is not necessarily in contradiction with the EFT approach since
short-distance physics included in the amplitude might be small in a
power counting sense. But if it is small it is not clear why it should
be included in the scattering amplitude.

An important observation in this spirit was made by van
Kolck~\cite{private} and KSW~\cite{Ka98a}. Given the experimentally
established hierarchy of scales $a\gg r_e \sim 1/M_\pi$, what the
effective theory should be reproducing is

\begin{equation}
T(k)=-\frac{{4\pi}}{M}
\frac{1}{({1}/{a}+ik)}
\left[1+ \frac{{r_e}/2}{({1}/{a}+ik)}{k^2} +O({k^4})\right],
\label{eq:cutoff3}
\end{equation}
and not necessarily the full effective range expansion of
Eq.~(\ref{eq:cutoff2}). This form of the scattering amplitude can be
reproduced in a scheme independent way by summing the $C$ operator of
Eq.~(\ref{eq:lag}) to all orders and treating all higher order
derivative operators as perturbations. Say $\aleph$ represents the
long-distance nonperturbative scale~\cite{private}. If $\Lambda$
represents the scale of short distance physics, then the effective
expansion parameter is $\aleph/{\Lambda}\sim 1/{a{M_\pi}}$.  Summing
to all orders in $1/{a{M_\pi}}$ gives the effective range expansion,
or equivalently, the transvestite.

A scheme in which this power counting is manifest was found by KSW in
Ref.~\cite{Ka98a}, which gives an elegant renormalization group
analysis of the coefficients in the EFT. The regularization and
renormalization scheme in which the power counting is manifest is $DR$
with power divergence subtraction ($PDS$).  As opposed to $MS$, in
which counterterms are added which subtract the poles in three space
dimensions, in $PDS$ the poles in two space dimensions are also
subtracted by counterterms. This scale-dependent scheme is similar to
performing a momentum subtraction at $p^2 =-{\mu^2}$~\cite{gegel}. In
this scheme, the fine-tuning in the underlying theory which gives rise
to a large scattering length is identified with a single operator in
the lagrangian, the $C$ operator of Eq.~(\ref{eq:lag}).

%%%%%%%%%%%%%%%%%%%%%%%%%%%%%%%%%%%%%%%%%%%%%%%%%%%%%%%%%%%%%%%%%%%%%%%
\section{The three-body force}

One of the most important results that has emerged from EFT in nuclear
physics is due to Bedaque, van Kolck and Hammer~\cite{Bd97a,Bd97b}.
These authors consider N-deuteron scattering.  There are two channels,
a quartet of total spin $J=3/2$ and a doublet of $J=1/2$. The leading
interactions involve two-body interactions whose low-energy parameters
have been fit to NN scattering. Recall that this is EFT at its best;
parameters fit to one process predict an independent process.  The
two-body interactions are accounted for using transvestite fields and
iterated using a Fadeev equation. This is not strictly systematic in
the sense of $\aleph/PDS$ power counting; however, including some of
the higher order terms in $\aleph/\Lambda$ via the transvestite does
not make the results any less accurate. Specifically, the transvestite
should be considered accurate to second order in the $\aleph$/PDS
power counting scheme.  Only the transvestite with spin one, isospin
zero contributes to the quartet scattering length, giving a
theoretical prediction of ${^4a}=6.33 \, {\rm fm}$ as compared to the
experimental value of ${^4a}=6.35\pm 0.02 \, {\rm fm}$.

One might wonder about the doublet channel in N-deuteron scattering.
Unlike the quartet channel, the scattering length in this channel is
not well described in the EFT because the absence of Pauli blocking
(which is present in the quartet channel) renders physics at short
distances potentially relevant to long distance observables.  Bedaque
and van Kolck have pointed out that the problem might be remedied by
inclusion of a 3-body contact interaction in the EFT which
``summarizes'' the effects of this short-distance physics. 

%%%%%%%%%%%%%%%%%%%%%%%%%%%%%%%%%%%%%%%%%%%%%%%%%%%%%%%%%%%%%%%%%%%%%%%
\section{The role of the pion -- a challenge for nuclear theorists}

It is desirable to push the short distance cut-off of the $NN$ EFT to
as high a momentum scale as possible. In a realistic EFT of $NN$
scattering it is important to include the pion.  The lightness of the
pion in itself guarantees that it should play a fundamental role in
nuclear physics.  However, it is the fact that chiral symmetry is
spontaneously broken --implying a light pion interacting weakly at low
energies-- that allows pion effects to be included in an EFT
description.

KSW have pointed out that Weinberg's power counting arguments are
problematic when computations are performed using dimensional
regularization~\cite{Ka96}. The fundamental problem is that pion
exchange effects in the ${}^3S_1-{}^3D_1$ channel that are leading
order in Weinberg's power counting require counterterms at all orders
in the momentum expansion, suggesting that Weinberg's power counting
scheme is not consistent.

Given the pathologies of the nonperturbative pion, KSW have proposed a
radical power counting scheme which fuses the $\aleph/PDS$ power
counting of the pionless effective theory with a perturbative
pion~\cite{Ka98a}. To date, phase shifts in the ${}^1S_0$ and
${}^3S_1-{}^3D_1$ channels have been computed in this scheme at
next-to-leading order. The ${}^3S_1-{}^3D_1$ mixing parameter
$\epsilon_1$ is a prediction at this order. Agreement with experiment
is reasonable. Moreover, KSW have calculated the electromagnetic form
factors of the deuteron at next-to-leading order using the parameters
fit to scattering data and have found good agreement with
experiment~\cite{Ka98b}.  This is a true test of the EFT.

The idea of a nuclear force with a perturbative pion is anathema to
most nuclear physicists. However, given that a consistent power
counting scheme has been proposed and nontrivial calculations have
been performed with good experimental agreement, it would seem
incumbent on traditionalists to propose low-energy observables whose
description requires a nonperturbative pion.

%%%%%%%%%%%%%%%%%%%%%%%%%%%%%%%%%%%%%%%%%%%%%%%%%%%%%%%%%%%%%%%%%%%%%%%
\section{Conclusion}

There has been remarkable progress made in the last few years in
developing systematic power counting technology for scattering
processes involving more than a single nucleon.  A new power counting
scheme, which is consistent in the sense of renormalization, has
emerged to challenge the original Weinberg power counting proposal.

Is Weinberg power counting wrong?  Is a nonperturbative pion truly
incompatible with EFT ideas?  The work of
Refs.~\cite{Or96,Sc97,Le97,park,steele} using cut-off EFT suggests
otherwise.  These numerical analyses include nonperturbative pions and
yet exhibit universal low-energy behavior: low-energy physics is
insensitive to the specific choice of regulator. One way of gaining
insight into this issue might be to unitarily transform the effective
potential (which is unobservable) to a new effective potential which
by construction involves only momenta less than a fixed
value~\cite{ulf}. In my view, understanding why EFT with a
nonperturbative pion works in spite of the failure implied by
dimensional regularization is an important issue, and not purely
academic. In losing Weinberg power counting we lose his beautiful
explanation of the hierarchy of n-body forces, which evidently has no
explanation in the new power counting scheme.

Be that as it may, it is clear that Kaplan, Savage and Wise have
introduced a consistent power counting scheme which is economical in
the sense that it appears to include only minimal short distance
physics and not the barrage of short distance physics which is
inherent to any exact solution of the Schr\"odinger equation.

%%%%%%%%%%%%%%%%%%%%%%%%%%%%%%%%%%%%%%%%%%%%%%%%%%%%%%%%%%%%%%%%%%%%%%%
\vspace{0.05in}
\noindent This work was supported by the U.S. Department of Energy grant
DE-FG02-93ER-40762. I thank Tom Cohen, Dan Phillips and Bira van Kolck
for valuable conversations.

\bibliographystyle{unsrt}

\begin{thebibliography}{9}       % For 0-9 references
% \begin{thebibliography}{99}      % For 10-99 references
%\bibitem{ward50}  J. C. Ward, {\it Phys.}\ {\it Rev.}\
% {\bf 78}, 182 (1950).

\bibitem{Be98a}
S.R.~Beane, T.D.~Cohen and D.R.~Phillips,
{\it Nucl.}\ {\it Phys.}\ {\bf A632}, 445  (1998), {\tt hep-th/9709062}.

\bibitem{Ka95}
D.B.~Kaplan, {\tt nucl-th/9506035}.

\bibitem{mano}
A.V.~Manohar, {\tt hep-ph/9506222}.

\bibitem{We90}
S.~Weinberg, {\it Phys.}\ {\it Lett.}\ {\bf B251},  288  (1990).

\bibitem{We91}
S.~Weinberg, {\it Nucl.}\ {\it Phys.}\ {\bf B363},  3  (1991).

\bibitem{Or96}
C.~Ord\~on\'ez, L.~Ray, and U.~van Kolck, 
{\it Phys.}\ {\it Rev.}\ {\bf C53}, 2086 (1996), {\tt nucl-th/9511380}.

\bibitem{Ka96}
D.B.~Kaplan, M.~Savage, and M.B.~Wise, 
{\it Nucl.}\ {\it Phys.}\ {\bf B478}, 629 (1996), {\tt nucl-th/9605002}.

\bibitem{Sc97}
K.A.~Scaldeferri, D.R.~Phillips, C.-W.~Kao, and T.D.~Cohen, 
{\it Phys.}\ {\it Rev.}\ {\bf C56}, 1 (1997), {\tt nucl-th/9610049}.

\bibitem{Le97}
G.P.~Lepage, {\tt nucl-th/9706029}.

\bibitem{park}
T-S.~Park {\it et al}, {\tt hep-ph/9711463}. 

\bibitem{Ubi94}
U.~van Kolck, 
{\it Phys.}\ {\it Rev.}\ {\bf C49}, 2932 (1994).

\bibitem{Fr96}
J.~L. Friar, {\tt nucl-th/9601012}.

\bibitem{St93}
V.G.J.~Stoks, R.A.M.~Klomp, M.C.M.~Rentmeester, and J.J.~de Swart, 
{\it Phys.}\ {\it Rev.}\ {\bf C48}, 792 (1993).

\bibitem{Ka97}
D.B.~Kaplan, {\it Nucl.}\ {\it Phys.}\ {\bf B494}, 471 (1997), 
{\tt nucl-th/9610052}.

\bibitem{tom}
T.D.~Cohen, {\it Phys.}\ {\it Rev.}\ {\bf C55}, 67 (1997), 
{\tt nucl-th/96006044}.

\bibitem{PC97}
D.R.~Phillips and T.D.~Cohen, 
{\it Phys.}\ {\it Lett.}\ {\bf B390}, 7 (1997), 
{\tt nucl-th/9607048}.

\bibitem{Ri97}
K.G.~Richardson, M.C.~Birse, and J.A.~McGovern, {\tt hep-ph/9708435}.

\bibitem{private}
U.~van Kolck, {\it private communication}; {\tt hep-ph/9711222}.

\bibitem{Ka98a}
D.B.~Kaplan, M.~Savage, and M.B.~Wise, {\tt nucl-th/9801034};
{\tt nucl-th/9802075}.

\bibitem{gegel}
J.~Gegelia, {\tt nucl-th/9802038}.

\bibitem{Bd97a}
P.F.~Bedaque and U.~van Kolck, 
{\it Phys.}\ {\it Lett.}\ {\bf B428}, 221 (1998), {\tt nucl-th/9710073}.

\bibitem{Bd97b}
P.F.~Bedaque, H.W.~Hammer and U.~van Kolck, {\tt nucl-th/9802057}.

\bibitem{Ka98b}
D.B.~Kaplan, M.~Savage, and M.B.~Wise, {\tt nucl-th/9804032}.

\bibitem{steele} J.V.~Steele and R.J.~Furnstahl, {\tt nucl-th/9802069}. 

\bibitem{ulf}
E.~Epelbaoum, W.~Gl{\"o}ckle and Ulf-G. Mei{\ss}ner,
{\tt nucl-th/9804005}.

\end{thebibliography}

\end{document}